\documentclass{llncs}
\usepackage{alltt}
\usepackage{listings}
\usepackage{epsfig}

\lstdefinelanguage{TF}
 {morekeywords=[1]{string, bytes, item, items, att, atts, Part, Token, Pol, List, URI, Addr, Polmap, Config, Secr, PrincipalSet, Link},
  morekeywords=[2]{new, if, then, else, in, out, begin, end, let, filter, done},
  morekeywords=[3]{private, channel, predicate, process, constructor, destructor, with, name, correspondence, secret, import, simulate},%
  morekeywords=[4]{form,proc,term,sort,ide,tm},
  sensitive=true,%
  morecomment=[l]{//*},
  morecomment=[s]{/*}{*/},
  morecomment=[s]{<!--}{-->},
  literate={->}{$\,\rightarrow\,$}3{!}{{\bfseries!}}2{|}{\boldmath$\mid$}3,
  morestring=[b]"}

\def\lstTF{
\lstset{language=TF,
        basicstyle=\sffamily\normalsize,%
        columns=[l]fullflexible,
        texcl=true,
        mathescape=true,
        xleftmargin=2pt,
        keywordstyle=[1]{\bfseries},
        keywordstyle=[2]{\bfseries},
        keywordstyle=[3]{\bfseries},
        keywordstyle=[4]{\rmfamily\itshape},
        rangeprefix=//---\ ,
        includerangemarker=false,
        stringstyle=\ttfamily,
        commentstyle=\rmfamily,
        moredelim=*[s][\ttfamily]{<}{>},
        breaklines=false}}

\lstTF

\usepackage{url}
\let\slash=/ \catcode`\/=\active \def/{\discretionary{\slash}{}{\slash}}

\def\und{\leavevmode{\kern0.03em\vbox{\hrule width0.47em}\kern0.03em}}

\newcommand{\hbra}{
\hbox to .995 \textwidth{\vrule width0.3mm height 1.8mm depth-0.3mm
                    \leaders\hrule height1.8mm depth-1.5mm\hfill
                    \vrule width0.3mm height 1.8mm depth-0.3mm}}
\newcommand{\hket}{
\hbox to .995 \textwidth{\vrule width0.3mm height1.5mm
                    \leaders\hrule height0.3mm\hfill
                    \vrule width0.3mm height1.5mm}}

\newif{\ifMarginalComments}
 \MarginalCommentsfalse  %

\newcommand{\ADG}[1]{%
  \ifMarginalComments{\marginpar{\tiny\bf[ADG says: #1]}} \else {} \fi}
\newcommand{\adg}[1]{%
  \ifMarginalComments{\marginpar{\tiny[adg--V.Next: #1]}} \else {} \fi}
\newcommand{\CF}[1]{%
  \ifMarginalComments{\marginpar{\tiny\bf[CF says: #1]}} \else {} \fi}

\newcommand{\cf}[1]{%
  \ifMarginalComments{\marginpar{\tiny[cf--V.Next: #1]}} \else {} \fi}

\newcommand{\comment}[1]{}

\newcommand{\wssec}{WS-Secu\-rity}
\newcommand{\picalc}{pi calculus}

\title{TulaFale: A Security Tool for Web Services}
\author{Karthikeyan Bhargavan \and C{\'e}dric Fournet \and \\ Andrew D. Gordon \and
  Riccardo Pucella}
\institute{Microsoft Research}

\begin{document}

\maketitle

\begin{abstract}
Web services security specifications are typically expressed as a mixture
of XML schemas, example messages, and narrative explanations.
We propose a new specification language for writing complementary
machine-checkable descriptions of SOAP-based security protocols and their
properties.
Our TulaFale language is based on the {\picalc} (for writing
collections of SOAP processors running in parallel), plus XML syntax
(to express SOAP messaging), logical predicates (to construct and
filter SOAP messages), and correspondence assertions (to specify
authentication goals of protocols).
Our implementation compiles TulaFale into the applied {\picalc}, and then
runs Blanchet's resolution-based protocol verifier.
Hence, we can automatically verify authentication properties of SOAP
protocols.
\end{abstract}

\section{Verifying Web Services Security}
\label{sec:intro}

Web services are a wide-area distributed systems technology,
based on asynchronous exchanges of XML messages conforming to the
SOAP message format~\cite{soap11,soap12}.
The WS-Security standard~\cite{ws-security-2004} describes how to sign and
encrypt portions of SOAP messages, so as to achieve end-to-end security.
This paper introduces TulaFale, a new language for defining and
automatically verifying models of SOAP-based cryptographic protocols,
and illustrates its usage for a typical request/response protocol:
we sketch the protocol, describe potential attacks, and then give a
detailed description of how to define and check the request and
response messages in TulaFale.
\CF{pls recheck first paragraph}

\subsection{Web Services}
A basic motivation for web services is to support programmatic access to
web data.
The HTML returned by a typical website is a mixture of data and presentational
markup, well suited for human browsing, but the presence of markup makes HTML
a messy and brittle format for data processing.
In contrast, the XML returned by a web service is just the data, with
some clearly distinguished metadata, well suited for programmatic access.
For example, search engines export web services for programmatic web search,
and e-commerce sites export web services to allow affiliated websites direct
access to their databases.

Generally, a broad range of applications for web services is emerging, from
the well-established use of SOAP as a platform and vendor neutral middleware
within a single organisation, to the proposed use of SOAP for device-to-device
interaction~\cite{WS-discovery}.

In the beginning, ``SOAP'' stood for ``Simple Object Access Protocol'', and was
intended to implement ``RPC using XML over
HTTP''~\cite{Winer98:RPCoverHTTPviaXML,Winer99:DavesHistoryofSOAP,Box01:ABriefHistoryofSOAP}.
HTTP facilitates interoperability between geographically distant machines and
between those in protection domains separated by corporate firewalls that block
many other transports.
XML facilitates interoperability between different suppliers' implementations,
unlike various binary formats of previous RPC technologies.
Still, web services technology should not be misconstrued as HTTP-specific RPC
for distributed objects~\cite{Vogels03:WebServicesAreNotDistributedObjects}.
HTTP is certainly at present the most common transport protocol, but the SOAP
format is independent of HTTP, and some web services use other
transports such as TCP or SMTP~\cite{soapmail}.
The design goals of SOAP/1.1~\cite{soap11} explicitly preclude object-oriented
features such as object activation and distributed garbage collection;
by version~1.2~\cite{soap12}, ``SOAP'' is a pure name, not an acronym.
The primitive message pattern in SOAP is a single one-way message that may be
processed by zero or more intermediaries between two end-points;
RPC is a derived message pattern built from a request and a response.
In brief, SOAP is not tied to objects, and web services are not tied to the web.
Still, our running example is an RPC over HTTP, which still appears to be the
common case.

\subsection{Securing Web Services with Cryptographic Protocols}

Web services specifications support SOAP-level security via
a syntax for embedding cryptographic materials in SOAP messages.
To meet their security goals, web services and their clients
can construct and check \emph{security headers} in messages, according
to the {\wssec} format
\cite{security-white-paper,ws-security-2004}.
{\wssec} can provide message confidentiality and authentication
independently of the underlying transport, using, for instance, secure
hash functions, shared-key encryption, or public-key cryptography.
{\wssec} has several advantages compared to using a secure transport
such as SSL, including scalability, flexibility, transparency to
intermediaries such as firewalls, and support for non-repudiation.
Significantly, though, {\wssec} does not itself prescribe a particular
security protocol: each application must determine its security goals, and
process security headers accordingly.

Web services may be vulnerable to many of the well-documented classes
of attack on ordinary
websites~\cite{SS02:HackingWebApplicationsExposed,HL03:WritingSecureCode}.
Moreover, unlike typical websites, web services relying on SOAP-based
cryptographic protocols may additionally be vulnerable to a new class of
\emph{XML rewriting attacks}: a range of attacks in which an attacker may
record, modify, replay, and redirect SOAP messages, but
without breaking the underlying cryptographic algorithms.
Flexibility comes at a price in terms of security, and it is surprisingly easy
to misinterpret the guarantees actually obtained from processing security
headers.
XML is hence a new setting for an old problem going back to Needham and
Schroeder's pioneering work on authentication protocols;
SOAP security protocols should be judged safe, or not, with respect to an
attacker who is able to ``interpose a computer on all communication paths, and
thus can alter or copy parts of messages, replay messages, or emit false
material''~\cite{NS78:EncryptionForAuthentication}.
XML rewriting attacks are included in the WS--I threat
model~\cite{wsiSecurityScenarios}.
We have found a variety of replay and impersonation attacks in practice.

\subsection{Formalisms and Tools for Cryptographic Protocols}

The use of formal methods to analyze cryptographic protocols and their
vulnerabilities begin with work by Dolev and
Yao~\cite{DY83:SecurityOfPublicKeyProtocols}.
In the past few years there has been intense research on the Dolev-Yao model,
leading to the development of numerous formalisms and tools.

TulaFale builds on the line of research using the {\picalc}.
The {\picalc}~\cite{Milner99:pi} is a general theory of interaction between
concurrent processes.
Several variants of the {\picalc}, including spi~\cite{AG99:spi},
and a generalization, applied pi~\cite{AbadiFournet:applied-pi-popl01},
have been used to formalize and prove properties of cryptographic protocols.
A range of compositional reasoning techniques is available for
proving protocol properties, but proofs typically require human skill
and determination.
Recently, however, Blanchet~\cite{BlanchetCSFW01,BlanchetSAS02} has proposed a
range of automatic techniques, embodied in his theorem prover ProVerif, for
checking certain secrecy and authentication properties of the applied
{\picalc}.
ProVerif works by compiling the {\picalc} to Horn clauses and then running
resolution-based algorithms.

\subsection{TulaFale: A Security Tool for Web Services}

TulaFale is a new scripting language for specifying SOAP security
protocols, and verifying the absence of XML rewriting attacks:
\begin{quote}
TulaFale = processes + XML + predicates + assertions
\end{quote}
The {\picalc} is the core of TulaFale, and allows us to describe SOAP
processors, such as clients and servers, as communicating processes.
We extend the {\picalc} with a syntax for XML plus symbolic
cryptographic operations; hence, we can directly express SOAP
messaging with {\wssec} headers.
We declaratively specify the construction and checking of SOAP
messages using Prolog-style predicates; hence, we can describe the
operational details of SOAP processing.
Independently, we specify security goals using various assertions, such
as correspondences for message authentication and correlation.

It is important that TulaFale can express the detailed structure of XML
signatures and encryption so as to catch low-level attacks on this structure,
such as copying part of an XML signature into another; more abstract
representations of message formats, typical in the study of the Dolev-Yao
model and used for instance in previous work on SOAP authentication
protocols~\cite{GP02:ValididatingWSSecurityAbstraction}, are insensitive to
such attacks.

Our methodology when developing TulaFale has been to study particular web
services implementations, and to develop TulaFale scripts modelling their
security aspects.
Our experiments have been based on the WSE development kit~\cite{wse02},
a particular implementation of WS-Security and related specifications.
We have implemented the running example protocol of this paper using WSE, and
checked that the SOAP messages specified in our script faithfully reflect the
SOAP messages observed in this implementation.
For a discussion of the implementation of related protocols, including
logs of SOAP messages, see the technical report version of an earlier
paper~\cite{BFG04:SemanticsForWebServicesAuthentication}.

\begin{figure}[t]
    \centerline{\includegraphics[width=4.0in]{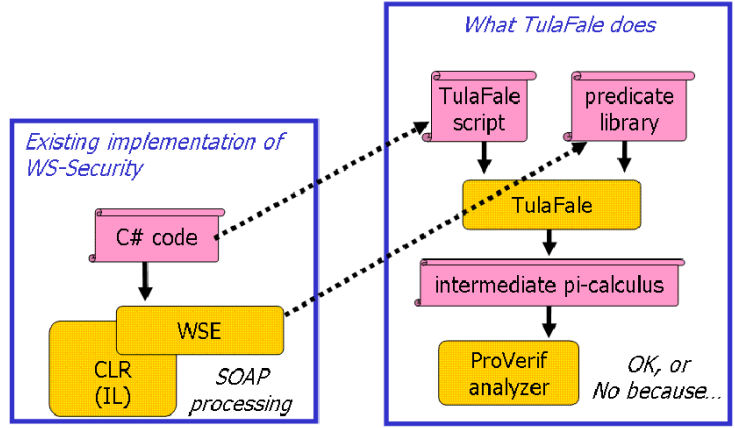}}
    \caption{Modelling WS-Security protocols with TulaFale}
    \label{fig:tulafale}
\end{figure}

Fig.~\ref{fig:tulafale} illustrates our methodology.
On the left, we have the user-supplied code for implementing a web services
protocol, such as the one of this paper, on top of the WSE library.
On the right, we have the TulaFale script modelling the user-supplied code, together
with some library predicates modelling operations performed by WSE.
Also on the right, we have the TulaFale tool, which compiles its input scripts
into the pure applied {\picalc} to be analyzed via ProVerif.

TulaFale is a direct implementation of the {\picalc} described in a
previous formal semantics of web services
authentication~\cite{BFG04:SemanticsForWebServicesAuthentication}.
The original contribution of this paper is to present a concrete language
design, to report an implementation of automatic verification of assertions in
TulaFale scripts using Blanchet's ProVerif, and to develop a substantial
example.

Section~\ref{sec:example}
informally introduces a simple request/response protocol and its
security goals: authentication and correlation of the two messages.
Section~\ref{sec:tulafale-basics} presents TulaFale syntax for XML with
symbolic cryptography and for predicates, and as a source of examples, explains
a library of TulaFale predicates for constructing and checking SOAP messages.
Section~\ref{sec:modelling-envelopes} describes  predicates
specific to the messages of our request/response protocol.
Section~\ref{sec:proc-assertions-TulaFale} introduces processes and
security assertions in TulaFale, and outlines their implementation via
ProVerif.
Section~\ref{sec:modelling-protocol} describes processes and
predicates specific to our protocol, and shows how to verify its
security goals.
Finally, Section~\ref{sec:conc} concludes.

\section{A Simple Request/Response Protocol}
\label{sec:example}

We consider a simple SOAP-based request/response protocol, of the kind easily
implemented using WSE to make an RPC to a web service.
Our security goals are simply message authentication and correlation.
To achieve these goals, the request includes a \emph{username token} identifying a
particular user and a \emph{signature token} signed by a key derived
from user's password; conversely, the response includes a signature
token signed by the server's public key.
Moreover, to preserve the confidentiality of the user's password from
dictionary attacks, the username token in the request message is encrypted
with the server's public key.
(For simplicity, we are not concerned here with any secrecy
properties, such as confidentiality of the actual message bodies, and
we do not model any authorization policies.)

In the remainder of this section, we present a detailed but informal
specification of our intended protocol, and consider some variations subject
to XML rewriting attacks.
Our protocol involves the following principals:
\begin{itemize}
\item A single certification authority (CA) issuing X.509 public-key
  certificates for services, signed with the CA's private key.
\item Two servers, each equipped with a public key certified by the CA and exporting
  an arbitrary number of web services.
\item Multiple clients, acting on behalf of human users.
\end{itemize}

Trust between principals is modelled as a database associating
passwords to authorized user names, accessible from clients and
servers.
Our threat model features an active attacker in control of the
network, in possession of all public keys and user names, but not in
possession of any of the following:
\begin{enumerate}
\item The private key of the CA.
\item The private key of any public key certified by the CA.
\item The password of any user in the database.
\end{enumerate}
The second and third points essentially rule out ``insider attacks''; we are
assuming that the clients, servers, and CA belong to a single close-knit
institution.
It is easy to extend our model to study the impact of insider attacks,
and indeed to allow more than two servers, but we
omit the details in this expository example.

\begin{figure}[t]
    \centerline{\includegraphics[width=4.0in]{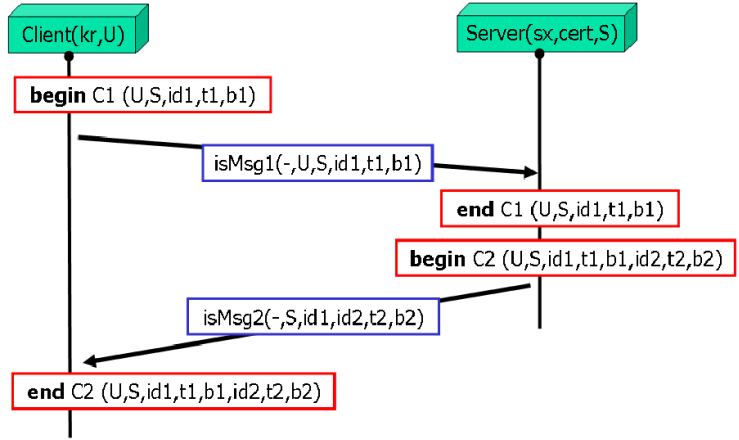}}
    \caption{An intended run of a client and server}
    \label{fig:intended-run}
\end{figure}
Fig.~\ref{fig:intended-run} shows an intended run of the protocol between
a client and server.
\begin{itemize}
\item
The principal \lstinline{Client(kr,U)} acts on behalf of a user
identified by \lstinline{U} (an XML encoding of the username and password).
The parameter \lstinline{kr} is the public key of the CA, needed by the
client to check the validity of public key certificates.
\item
The principal \lstinline{Server(sx,cert,S)} implements a service identified by
\lstinline{S} (an XML encoding of a URL address, a SOAP action, and the
subject name appearing on the service's certificate).
The parameter \lstinline{sx} is the server's private signing key,
while \lstinline{cert} is its public certificate.
\item
The client sends a request message satisfying
\lstinline{isMsg1(-,U,S,id1,t1,b1)}, which we define later to mean the message
has body \lstinline{b1}, timestamp \lstinline{t1}, and message identifier
\lstinline{id1}, is addressed to a web service \lstinline{S}, and has a
\lstinline{<Security>} header containing a token identifying \lstinline{U} and
encrypted with \lstinline{S}'s public key, and a signature of \lstinline{S},
\lstinline{id1}, \lstinline{t1}, and \lstinline{b1} by \lstinline{U}.
\item
The server sends a response message satisfying
\lstinline{isMsg2(-,S,id1,id2,t2,b2)}, which we define later to mean the
message has body \lstinline{b2}, timestamp \lstinline{t2}, and message
identifier \lstinline{id2}, is sent from \lstinline{S}, and has a
\lstinline{<Security>} header containing \lstinline{S}'s certificate
\lstinline{cert} and a signature of \lstinline{id1}, \lstinline{id2},
\lstinline{t2}, and \lstinline{b2} by \lstinline{S}.
\item
The client and server enact begin- and end-events labelled
\lstinline{C1(U,S,id1,t1,b1)}
to record the data agreed after receipt of the first message.
Similarly, the begin- and end-events labelled
\lstinline{C2(U,S,id1,t1,b1,id2,t2,b2)}
record the data agreed after both messages are received.
Each begin-event marks an intention to send data.
Each end-event marks apparently successful agreement on data.
\end{itemize}

The begin- and end-events define our authentication and correlation
goals: for every end-event with a particular label, there is a
preceding begin-event with the same label in any run of the system,
even in the presence of an active attacker.
\CF{rewritten; pls check}%
Such goals are known as one-to-many correspondences~\cite{WL93:SemanticModel} or
non-injective agreements~\cite{Lowe97:HierarchyAuthenticationSpecs}.
The \lstinline{C1} events specify authentication of the request,
while the \lstinline{C2} events specify authentication of the
response. By including data from the request,
\lstinline{C2} events also specify correlation of the request and response.

Like most message sequence notations, Fig.~\ref{fig:intended-run} simply
illustrates a typical protocol run, and is not in itself an adequate
specification.
In Sections \ref{sec:modelling-envelopes} and~\ref{sec:modelling-protocol} we present a formal specification in
TulaFale:
\cf{These are not principals, and not quite roles.}%
we define the principals \lstinline{Client(kr,U)}
and \lstinline{Server(sx,cert,S)} as parametric processes,
and we define the checks \lstinline{isMsg1} and \lstinline{isMsg2}
as predicates on our model of XML with symbolic cryptography.
The formal model clarifies the following points, which are left
implicit in the figure:
\begin{itemize}
\item
The client can arbitrarily choose which service \lstinline{S} to call, and
which data \lstinline{t1} and \lstinline{b1} to send.
(In the formal model, we typically make such arbitrary choices by inputting the
data from the opponent.)
Conversely, the client must
generate a fresh identifier \lstinline{id1} for each request, or else it is
impossible to correlate the responses from two simultaneous requests to the
same service.
\item
Similarly, the server can arbitrarily choose the response data \lstinline{id2},
\lstinline{t2}, and \lstinline{b2}.
\end{itemize}

On the other hand, our formal model does not directly address replay protection.
To rule out direct replays of correctly signed messages, we would need to
specify that for each end-event there is a unique preceding begin-event with
the same label.
This is known as a one-to-one correspondence or injective agreement.
In practice, we can protect against direct replays using a cache of
recently received message identifiers and timestamps to ensure that no two
messages are accepted with the same identifier and timestamp.
Hence, if we can prove that the protocol establishes non-injective
agreement on data including the identifiers and timestamps, then,
given such replay protection, the protocol implementation also
establishes injective agreement.

\begin{figure}[t]
    \centerline{\includegraphics[width=4.0in]{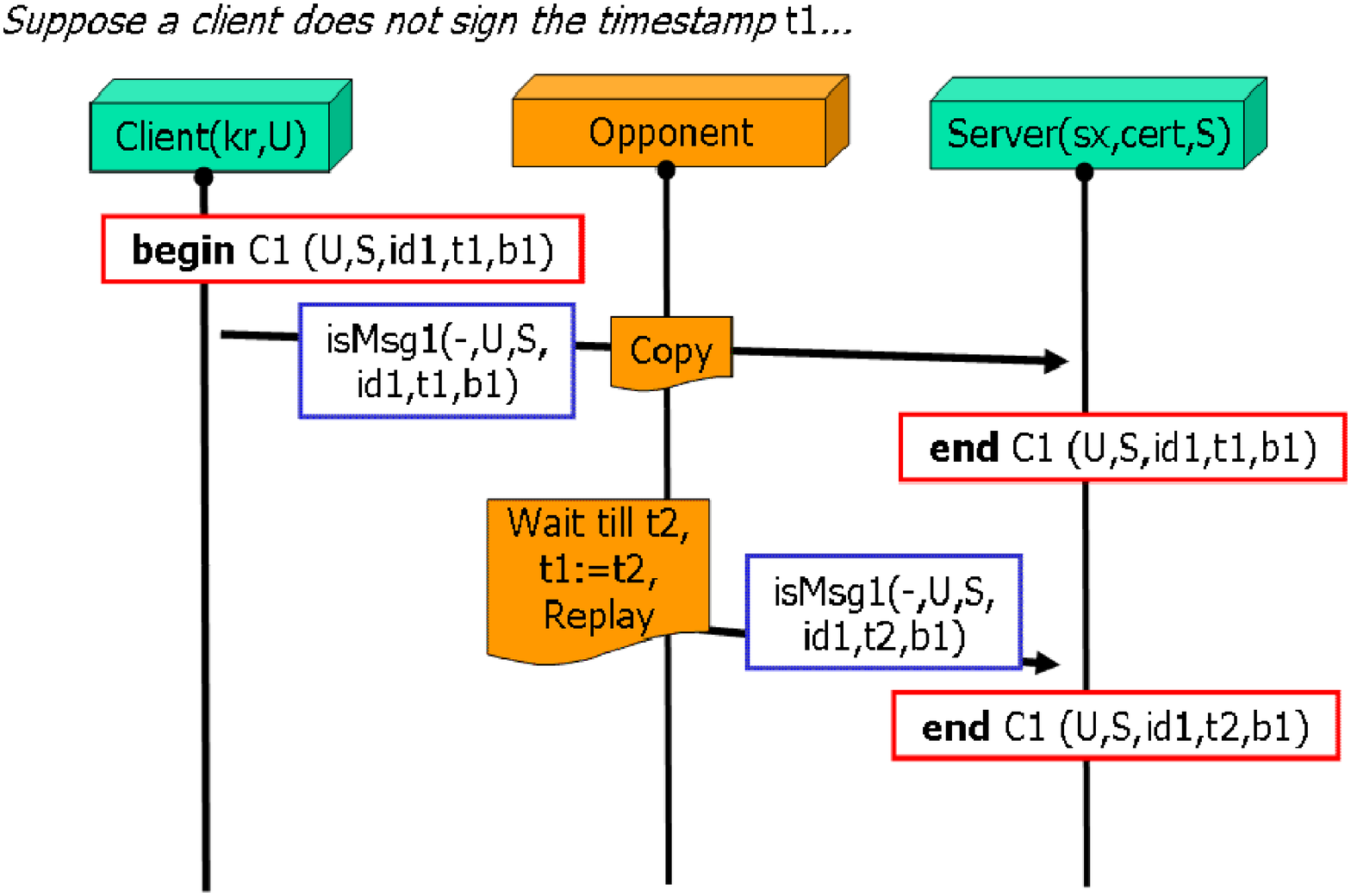}}
    \caption{A replay attack}
    \label{fig:replay-attack}
\end{figure}

\begin{figure}[t]
    \centerline{\includegraphics[width=4.0in]{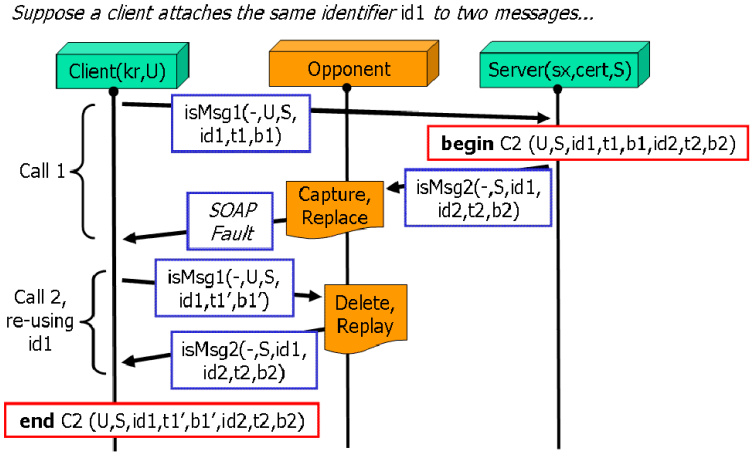}}
    \caption{A failure of message correlation}
    \label{fig:correlation-attack}
\end{figure}

We end this section by discussing some flawed variations of the protocol,
corresponding to actual flaws we have encountered in user code for web
services.
\begin{itemize}
\item
Suppose that the check \lstinline{isMsg1(-,U,S,id1,t1,b1)} only requires that
\lstinline{S}, \lstinline{id1}, and \lstinline{b1}, are signed by
\lstinline{U}, but not the timestamp \lstinline{t1}.
Replay protection based on the timestamp is now ineffective: the opponent can
record a message with timestamp \lstinline{t1}, wait until some time
\lstinline{t2} when the timestamp has expired, and the message
identifier \lstinline{id1} is no longer being cached,
rewrite the original message with timestamp \lstinline{t2}, and then replay
the message.
The resulting message satisfies \lstinline{isMsg1(-,U,S,id1,t2,b1)},
since \lstinline{t2} does not need to be signed,
and hence is accepted by the server.
Fig.~\ref{fig:replay-attack} shows the attack, and the resulting
failure of correspondence \lstinline{C1}.
\item
Suppose that a client re-uses the same message identifier in two
different calls to a web service; the opponent can manipulate messages so that
the client treats the response to the first call as if it were the response to
the second call.
Fig.~\ref{fig:correlation-attack} shows the attack.
The client sends a first request with body \lstinline{b1} and identifier
\lstinline{id1}.
The opponent intercepts the response with body \lstinline{b2}, and sends
a SOAP fault back to the client.
Subsequently, the client sends a second request with the same identifier
\lstinline{id1} as the first, and body \lstinline{b1'}.
The opponent can delete this request to prevent it reaching the service,
and then replay the original response.
The client now considers that \lstinline{b2} is the response to
\lstinline{b1'}, when in fact it is the response to \lstinline{b1}, perhaps
completely different.
Formally, this is a failure of correspondence \lstinline{C2}.
\item
Suppose that the server does not include the request identifier \lstinline{id1}
in the signature on the response message.  Then the opponent can mount a
similar correlation attack, breaking \lstinline{C2}---we omit the details. 
\end{itemize}

We can easily adapt our TulaFale script to model these variations in the
protocol. Our tool automatically and swiftly detects the errors,
and returns descriptions of the messages sent during the attacks.
These flaws in web services code are typical of errors
in cryptographic protocols historically.
The practical impact of these flaws is hard to assess, as they were found in
preliminary code, before deployment.
Still, it is prudent to eliminate these vulnerabilities, and
tools such as TulaFale can systematically rule them out.

\section{XML, Principals, and Cryptography in TulaFale}
\label{sec:tulafale-basics}

This section introduces the term and predicate language of TulaFale,
via a series of basic constructions needed for the example protocol of
Section~\ref{sec:example}.
Throughout the paper, for the sake of exposition, we elide some details of
SOAP envelopes, such as certain headers and attributes, that are unconnected
to security.

\subsection{XML Elements, Attributes, and Strings}

Here is a TulaFale term for a SOAP request, illustrating the format of the
first message in our example protocol:
\begin{lstlisting}
<Envelope>
  <Header>
    <To>uri</>
    <Action>ac</>
    <MessageId>id</>
    <Security>
      <Timestamp><Created>"2004-03-19T09:46:32Z"</></>
      utok
      sig
    </>
  </>
  <Body Id="1">request</>
</>
\end{lstlisting}

\adg{Policy: put angle brackets around element names}

Every SOAP message consists of an XML \lstinline@<Envelope>@ element, with two
children: an optional \lstinline@<Header>@ and a mandatory \lstinline@<Body>@.
In this example, the header has four children, and the body
has an \lstinline@Id@-attribute, the literal string \lstinline@"1"@.

We base TulaFale on a sorted (or typed) term algebra, built up from a set of
function symbols and variables.
The basic sorts for XML data include \lstinline{string} (for string literals), \lstinline{att} (for named attributes), and \lstinline{item} (either an element or a string).
Every element or attribute tag (such as \lstinline@Envelope@ or
\lstinline@Id@, for example) corresponds to a sorted function symbol in
the underlying algebra.

Although TulaFale syntax is close to the XML wire format, it is not identical.
We suppress all namespace information.
As previously mentioned, we omit closing element tags; for example, we write \lstinline@</>@ instead of
\lstinline@</Envelope>@.
Literal strings are always quoted, as in
\lstinline@<Created>"2004-03-19T09:46:32Z"</>@.
In the standard wire format, the double quotes would be omitted
when a string is an element body.
We use quotation to distinguish strings from term variables, such as the
variables \lstinline{uri}, \lstinline{ac}, \lstinline{id}, \lstinline{utok},
\lstinline{sig}, and \lstinline{request} in the example above.

\subsection{Symbolic Cryptography}

In TulaFale, we represent cryptographic algorithms symbolically, as function
symbols that act on a sort \lstinline{bytes} of byte arrays.
Each function is either a data constructor, with no accompanying rewrite rule,
or it is a destructor, equipped with a rewrite rule for testing or extracting
data from an application of a constructor.
For example, encryption functions are constructors, and decryption functions
are destructors.
This approach, the basis of the Dolev-Yao
model~\cite{DY83:SecurityOfPublicKeyProtocols}, assumes that the underlying
cryptography is \emph{perfect}, and can be faithfully reflected by abstract
equational properties of the functions.
It also abstracts some details, such as the lengths of strings and byte
arrays.
The TulaFale syntax for introducing constructors and destructors is based on
the syntax used by ProVerif.

For instance, we declare function symbols for RSA key generation, public-key
encryption, and private-key decryption using the following TulaFale declarations:
\begin{lstlisting}
constructor pk(bytes):bytes.
constructor rsa(bytes,bytes):bytes.
destructor decrsa(bytes,bytes):bytes with
  decrsa(s,rsa(pk(s),b)) = b.
\end{lstlisting}
The constructor \lstinline{pk} represents the relationship between private and
public keys (both byte arrays, of sort~\lstinline{bytes}); it takes a private key
and returns the corresponding public key.
There is no inverse or destructor, as we intend to represent a
one-way function: given only \lstinline{pk(s)} it is impossible to extract
\lstinline{s}.

The constructor \lstinline{rsa(k,x)} encrypts the data \lstinline{x:bytes} under
the public key~\lstinline{k}, producing an encrypted byte array.
The destructor \lstinline{decrsa(s,e)} uses the corresponding private key
\lstinline{s} to decrypt a  byte array generated by \lstinline{rsa(pk(s),x)}. 
The destructor definition expresses the decryption operation as a rewrite rule:
when an application of \lstinline{decrsa} in a term matches the left-hand side
of the rule, it may be replaced by the corresponding right-hand side. 

To declare RSA public-key signatures, we introduce another
constructor \lstinline{rsasha1(s,x)} that produces a
RSA signature of a cryptographic hash of data \lstinline{x} under the
private key \lstinline{s}:
\begin{lstlisting}
constructor rsasha1(bytes,bytes):bytes.
destructor checkrsasha1(bytes,bytes,bytes):bytes with 
  checkrsasha1(pk(s),x,rsasha1(s,x))=pk(s).
\end{lstlisting}
To check the validity of a signature \lstinline{sig} on \lstinline{x} using
a public key~\lstinline{k}, one can form the term \lstinline{checkrsasha1(k,x,sig)}
and compare it to \lstinline{k}.
If \lstinline{k} is a public key of the form~\lstinline{pk(s)} and \lstinline{sig} is
the result of signing \lstinline{x} under the corresponding private key \lstinline{s}, 
then this term rewrites to \lstinline{k}.

\adg{have changed ``root authority'' to ``CA'', since we are not modelling
  certification chains and their roots, just a single CA}

For the purposes of this paper, an X.509 certificate binds a key to a
subject name by embedding a digital signature generated from the
private key of some certifying authority (CA).
We declare X.509 certificates as follows:
\begin{lstlisting}
constructor x509(bytes,string,string,bytes):bytes.
destructor x509key(bytes):bytes with 
  x509key(x509(s,u,a,k))=k.
destructor x509user(bytes):string with
  x509user(x509(s,u,a,k))=u.
destructor x509alg(bytes):string with 
  x509alg(x509(s,u,a,k))=a.
destructor checkx509(bytes,bytes):bytes with 
  checkx509(x509(sr,u,a,k),pk(sr))=pk(sr).
\end{lstlisting}
The term \lstinline{x509(sr,u,a,k)} represents a certificate that binds the
subject name~\lstinline{u} to the public key \lstinline{k}, for use with the
signature algorithm \lstinline{a} (typically \lstinline{rsasha1}).
This certificate is signed by the CA with private key \lstinline{sr}. 
Given such a certificate, the destructors \lstinline{x509key}, \lstinline{x509user},
and \lstinline{x509alg} extract the three public fields of the certificate.
Much like \lstinline{checkrsasha1} for ordinary digital signatures, an
additional destructor \lstinline{checkx509} can be used to check the
authenticity of the embedded signature.

\subsection{XML Encryption and Decryption}

Next, we write logical predicates to construct and parse XML encrypted
under some known RSA public key. 
A predicate is written using a Prolog-like syntax;
it takes a tuple of terms and checks logical properties, 
such as whether two terms are equal or whether a term
has a specific format. 
It is useful to think of some of the terms given to the 
predicate as inputs and the others as outputs. 
Under this interpretation, the predicate computes output terms that satisfy
the logical properties by pattern-matching.

The predicate \lstinline{mkEncryptedData} takes a plaintext 
\lstinline{plain:item} and an RSA public encryption key \lstinline{ek:bytes}, and it
generates an XML element \lstinline{encrypted} containing
the XML encoding of \lstinline{plain} encrypted under \lstinline{ek}.
\begin{lstlisting}
predicate mkEncryptedData (encrypted:item,plain:item,ek:bytes) :-
  cipher = rsa(ek,c14n(plain)),
  encrypted = <EncryptedData>
                <CipherData>
                  <CipherValue>base64(cipher)</></></>.
\end{lstlisting}
The first binding in the predicate definition computes the encrypted byte
array, \lstinline{cipher}, using the \lstinline{rsa} 
encryption function applied to the key \lstinline{ek} and the plaintext \lstinline{plain}.
Since \lstinline{rsa} is only defined over byte arrays, 
\lstinline{plain:item} is first converted to \lstinline{bytes} using the (reversible) 
\lstinline{c14n} constructor. 
The second binding generates an XML element (\lstinline{<EncryptedData>})
containing the encrypted bytes. Since only strings or elements can be embedded into XML elements, 
the encrypted byte array, \lstinline{cipher}, is first converted to a string using the 
(reversible) \lstinline{base64} constructor.
\ADG{explain that c14n does canonicalization followed by byte encoding of XML}

In this paper, we use three transformation functions between sorts:
\lstinline{c14n} (with inverse \lstinline{ic14n}) transforms an \lstinline{item} to a
\lstinline{bytes}, \lstinline{base64} (with inverse \lstinline{ibase64}) transforms a
\lstinline{bytes} to a \lstinline{string}, and \lstinline{utf8} (with inverse
\lstinline{iutf8}) transforms a \lstinline{string} to a \lstinline{bytes}.  All three
functions have specific meanings in the context of XML transformations, but we
treat them simply as coercion functions between sorts.

To process a given element, \lstinline{<Foo>} say, we sometimes write
distinct predicates on the sending side and on the receiving side of the
protocol, respectively. By convention, to construct a \lstinline{<Foo>} element,
we write a predicate named \lstinline{mkFoo} whose first parameter is the element
being constructed; to parse and check it, we write a predicate named
\lstinline{isFoo}.

For \lstinline{<EncryptedData>}, for instance, the logical predicate
\lstinline{isEncryptedData} parses elements constructed by
\lstinline{mkEncryptedData}; it takes an element \lstinline{encrypted} and a
decryption key \lstinline{dk:bytes} and, if \lstinline{encrypted} is
an \lstinline{<EncryptedData>} element with some plaintext encrypted
under the corresponding encryption key \lstinline{pk(dk)}, it returns the
plaintext as \lstinline{plain}.
\begin{lstlisting}
predicate isEncryptedData (encrypted:item,plain:item,dk:bytes) :-
  encrypted = <EncryptedData>
                <CipherData>
                  <CipherValue>base64(cipher)</></></>,
  c14n(plain) = decrsa(dk,cipher).
\end{lstlisting}

Abstractly, this predicate reverses the computations performed by
\lstinline$mkEncryp$\-\lstinline$tedData$.
One difference, of course, is that while \lstinline{mkEncryptedData} is passed
the public encryption key \lstinline{ek}, the \lstinline{isEncryptedData} predicate
is instead passed the private key, \lstinline{dk}.
The first line matches \lstinline{encrypted} against a pattern
representing the \lstinline{<EncryptedData>} element, extracting the
encrypted byte array, \lstinline{cipher}.  
Relying on pattern-matching, 
the constructor \lstinline{base64} is implicitly inverted using its destructor
\lstinline{ibase64}.  The second line decrypts \lstinline{cipher} using the
decryption key \lstinline{dk}, and implicitly inverts \lstinline{c14n}
using its destructor \lstinline{ic14n} to compute \lstinline{plain}.

\subsection{Services and X509 Security Tokens}

We now implement processing for the service identifiers used
in Section~\ref{sec:example}.
We identify each web service by a structure consisting of a
\lstinline{<Service>} element containing \lstinline{<To>},
\lstinline{<Action>}, and \lstinline{<Subject>} elements.
For message routing, the web service is identified by the HTTP URL \lstinline{uri}
where it is located and the name of the action \lstinline{ac} to be invoked at
the URL.
(In SOAP, there may be several different actions available at the same \lstinline{uri}.)
The web service is then willing to accept any SOAP message with a \lstinline{<To>} header
containing \lstinline{uri} and an \lstinline{<Action>} header containing \lstinline{ac}.
Each service has a public key with which parts of requests may be encrypted,
and parts of responses signed.
The \lstinline{<Subject>} element contains the subject name bound to the server's
public key by the X.509 certificate issued by the CA.
For generality, we do not assume any relationship between the URL and subject
name of a service, although in practice the subject name might contain the
domain part of the URL.

The logical predicate \lstinline{isService} parses a \lstinline{service} element to
extract the \lstinline{<To>} field as \lstinline{uri}, the \lstinline{<Action>} field as
\lstinline{ac}, and the \lstinline{<Subject>} field as \lstinline{subj}:
\begin{lstlisting}
predicate isService(S:item,uri:item,ac:item,subj:string) :-
  S = <Service><To>uri</> <Action>ac</> <Subject>subj</></>.
\end{lstlisting}

We also define predicates to parse X.509 certificates and
to embed them in SOAP headers:
\begin{lstlisting}
predicate isX509Cert (xcert:bytes,kr:bytes,u:string,a:string,k:bytes) :-
  checkx509(xcert,kr) = kr,
  u = x509user(xcert),
  k = x509key(xcert),
  a = x509alg(xcert).

predicate isX509Token (tok:item,kr:bytes,u:string,a:string,k:bytes) :-
  tok = <BinarySecurityToken ValueType="X509v3">base64(xcert)</>,
  isX509Cert (xcert,kr,u,a,k).
\end{lstlisting}
The predicate \lstinline{isX509Cert} takes a byte array \lstinline{xcert} containing
an X.509 certificate, checks that it has been issued by a certifying authority with
public key \lstinline{kr}, and extracts the user name \lstinline{u}, 
its user public key \lstinline{k}, and its signing algorithm~\lstinline{a}.
In SOAP messages, certificates are carried in XML elements called
security tokens.  The predicate \lstinline{isX509Token} checks that an XML token
element contains a valid X.509 certificate and extracts the relevant
fields.

\subsection{Users and Username Security Tokens}

In our system descriptions, we identify each user by a \lstinline{<User>}
element that contains their username and password.
The predicate \lstinline{isUser} takes such an element, \lstinline{U}, and
extracts its embedded username \lstinline{u} and password \lstinline{pwd}.
\begin{lstlisting}%
predicate isUser (U:item,u:string,pwd:string) :-
  U = <User><Username>u</><Password>pwd</></>.
\end{lstlisting}

In SOAP messages, the username is represented by a 
\lstinline{Userna}\-\lstinline{meToken} that contains the \lstinline{<Username>} \lstinline{u},
a freshly generated nonce \lstinline{n}, and a timestamp~\lstinline{t}.
The predicate \lstinline{isUserTokenKey} takes such a token \lstinline{tok} 
and extracts \lstinline{u}, \lstinline{n}, \lstinline{t}, and 
then uses a user \lstinline{U} for \lstinline{u} to 
compute a key from \lstinline{pwd}, \lstinline{n}, and \lstinline{t}.
\begin{lstlisting}
predicate isUserTokenKey (tok:item,U:item,n:bytes,t:string,k:bytes) :-
  isUser(U,u,pwd),
  tok = <UsernameToken @ _>
           <Username>u</>
           <Nonce>base64(n)</>
           <Created>t</></>,
  k = psha1(pwd,concat(n,utf8(t))).
\end{lstlisting}
The first line parses \lstinline{U} to extract the username \lstinline{u} and
password \lstinline{pwd}.
The second line parses \lstinline{tok} to extract \lstinline{n} and \lstinline{t},
implicitly checking that the username~\lstinline{u} is the same.
In TulaFale, lists of terms are written as \lstinline!tm$_1$ $\dots$ tm$_m$ @ tm! with $m \geq 0$, 
where the terms \lstinline{tm$_1$}, $\dots$, 
\lstinline{tm$_m$} are the first $m$ members of the list, and
the optional term {\lstinline{tm}} is the rest of the list.
Here, the wildcard \lstinline{@ _} of the \lstinline{<UsernameToken>}
element matches the entire list of attributes.
The last line computes the key \lstinline{k} by applying the cryptographic
hash function \lstinline{psha1} to \lstinline{pwd}, \lstinline{n}, and
\lstinline{t} (converted to \lstinline{bytes}).
(This formula for \lstinline{k} is a slight simplification of the actual key
derivation algorithm used by WSE.)
The \lstinline{concat} function returns the concatenation of two byte arrays.

\subsection{Polyadic Signatures}

An XML digital signature consists of a list of references to the elements
being signed, together with a \emph{signature value} that binds hashes of
these elements using some signing secret.
The signature value can be computed using several different
cryptographic algorithms; in our example protocol, we rely on
\lstinline{hmacsha1} for user signatures and on \lstinline{rsasha1} for
service signatures.

The following predicates describe how to construct (\lstinline{mkSigVal})
and check (\lstinline{isSigVal}) 
the signature value~\lstinline{sv} of an XML element \lstinline{si},
signed using the algorithm \lstinline{a} with a key \lstinline{k}. 
Each of these predicates is defined by a couple of clauses,
representing symmetric and asymmetric signature algorithms.
When a predicate is defined by multiple clauses, they are interpreted as a
disjunction; that is, the predicate is satisfied if any one of its clauses is
satisfied.
\begin{lstlisting}
predicate mkSigVal (sv:bytes,si:item,k:bytes,a:string) :-
  a = "hmacsha1", sv = hmacsha1(k,c14n(si)).

predicate isSigVal (sv:bytes,si:item,k:bytes,a:string) :-
  a = "hmacsha1", sv = hmacsha1(k,c14n(si)).

predicate mkSigVal (sv:bytes,si:item,k:bytes,a:string) :-
  a = "rsasha1",  sv = rsasha1(k, c14n(si)).

predicate isSigVal (sv:bytes,si:item,p:bytes,a:string) :-
  a = "rsasha1",  p = checkrsasha1(p,c14n(si),sv).
\end{lstlisting}

The first clause of \lstinline{mkSigVal} takes an item \lstinline{si}
to be signed and a key~\lstinline{k} for the symmetric signing algorithm
\lstinline{hmacsha1}, and generates the signature value~\lstinline{sv}.
The first clause of \lstinline{isSigVal} reverses this computation,
taking \lstinline{sv}, \lstinline{si}, \lstinline{k}, and \lstinline{a = "hmacsha1"}
as input and checking that \lstinline{sv} is a valid signature value of
\lstinline{si} under~\lstinline{k}.
Since the algorithm is symmetric, the two clauses are identical.
The second clause of \lstinline{mkSigVal} computes the signature value using the
asymmetric \lstinline{rsasha1} algorithm, and the corresponding clause of
\lstinline{isSigVal} checks this signature value.
In contrast to the symmetric case, the two clauses rely on different
computations.

A complete XML signature for a SOAP message contains both the
signature value \lstinline{sv}, as detailed above, and an explicit
description of the message parts are used to generate
\lstinline{si}.
Each signed item is represented by a \lstinline{<Reference>} element.

The predicate \lstinline{mkRef} takes an item \lstinline{t} and generates a
\lstinline{<Reference>} element~\lstinline{r} by embedding a \lstinline{sha1}
hash of \lstinline{t}, with appropriate sort conversions.
Conversely, the predicate \lstinline{isRef} checks that \lstinline{r} is a
\lstinline{<Reference>} for \lstinline{t}.
\begin{lstlisting}
predicate mkRef(t:item,r:item) :-
  r = <Reference> 
        <Other></> <Other></> 
        <DigestValue> base64(sha1(c14n(t))) </> </>.

predicate isRef(t:item,r:item) :-
  r = <Reference> 
         _ _ 
        <DigestValue> base64(sha1(c14n(t))) </> </>.
\end{lstlisting}
(The XML constructed by \lstinline{mkRef} abstracts some of the detail that is
included in actual signatures, but that tends not to vary in practice; in
particular, we include \lstinline{<Other>} elements instead of the standard
\lstinline{<Transforms>} and \lstinline{<DigestMethod>} elements.  On the other
hand, the \lstinline{<DigestValue>} element is the part that depends on the
subject of the signature, and that is crucial for security, and we model this
element in detail.)

More generally, the predicate \lstinline{mkRefs(ts,rs)} constructs a list
\lstinline{ts} and from a list \lstinline{rs}, such that their members are pairwise
related by \lstinline{mkRef}.
Similarly, the predicate \lstinline{mkRefs(ts,rs)} checks that two given lists
are pairwise related by \lstinline{mkRef}.
We omit their definitions.

A \lstinline{<SignedInfo>} element is constructed from
\lstinline{<Reference>} elements for every signed element. A
\lstinline{<Signature>} element consists of a \lstinline{<SignedInfo>}
element \lstinline{si} and a \lstinline{<SignatureValue>} element
containing~\lstinline{sv}.
Finally, the following predicates define how signatures are constructed and
checked.
\begin{lstlisting}
predicate mkSigInfo (si:item,a:string,ts:item) :- 
  mkRefs(ts,rs),
  rs = <list>@ refs</>,
  si = <SignedInfo>
         <Other></> <SignatureMethod Algorithm=a> </> 
         @ refs </>.

predicate isSigInfo (si:item,a:string,ts:item) :-
  si = <SignedInfo> 
         _ <SignatureMethod Algorithm=a> </> 
         @ refs</>,
  rs = <list>@ refs</>,
  isRefs(ts,rs).

predicate mkSignature (sig:item,a:string,k:bytes,ts:item) :-
  mkSigInfo(si,a,ts),
  mkSigVal(sv,si,k,a),
  sig = <Signature> si <SignatureValue> base64(sv) </> </>.

predicate isSignature (sig:item,a:string,k:bytes,ts:item) :-
  sig = <Signature> si <SignatureValue> base64(sv) </>@ _</>,
  isSigInfo(si,a,ts),
  isSigVal(sv,si,k,a).
\end{lstlisting}

The predicate \lstinline{mkSigInfo} takes a list of items to be signed, 
embedded in a \lstinline{<list>} element \lstinline{ts}, and generates 
a list of references \lstinline{refs} for them, embedded in a
\lstinline{<list>} element \lstinline{rs}, which
are then embedded into \lstinline{si}.
The predicate \lstinline{isSigInfo} checks \lstinline{si} has been correctly
constructed from \lstinline{ts}.

The predicate \lstinline{mkSignature} constructs \lstinline{si} using
\lstinline{mkSigInfo}, generates the signature value \lstinline{sv} using
\lstinline{mkSigVal}, and puts them together in a \lstinline{<Signature>} element 
called \lstinline{sig}; \lstinline{isSignature} checks that a signature \lstinline{sig}
has been correctly generated from \lstinline{a}, \lstinline{k}, and \lstinline{ts}.

\section{Modelling SOAP Envelopes for our protocol}
\label{sec:modelling-envelopes}

Relying on the predicate definitions of Section~\ref{sec:tulafale-basics},
which reflect (parts of) the SOAP and {\wssec} specifications but do not
depend on the protocol, we now define custom ``top-level'' predicates to build
and check Messages 1 and~2 of our example protocol.

\subsection{Building and Checking Message~1}
\label{sec:message1}

Our goal \lstinline{C1} is to reach agreement on the data
\begin{quote}
\lstinline@(U,S,id1,t1,b1)@
\end{quote}
where
\begin{quote}
\lstinline@U@=\lstinline@<User><Username>u</><Password>pwd</></>@\\
\lstinline@S@=\lstinline@<Service><To>uri</><Action>ac</><Subject>subj</></>@
\end{quote}
after receiving and successfully checking Message~1.
To achieve this, the message includes a username token for \lstinline@U@,
encrypted with the public key of \lstinline@S@ (that is, one whose
certificate has the subject name \lstinline@subj@), and also includes a 
signature token, signing  (elements containing) \lstinline@uri@, \lstinline@ac@,
\lstinline@id1@, \lstinline@t1@, \lstinline@b1@, and the encrypted
username token, signed with the key derived from the username token.

We begin with a predicate setting the structure of the first envelope:
\begin{lstlisting}
predicate env1(msg1:item,uri:item,ac:item,id1:string,t1:string,
               eutok:item,sig1:item,b1:item) :-
  msg1 =
    <Envelope>
      <Header>
        <To>uri</>
        <Action>ac</>
        <MessageId>id1</>
        <Security>
          <Timestamp><Created>t1</></>
          eutok
          sig1</></>
      <Body>b1</></>.
\end{lstlisting}

On the client side, we use a predicate \lstinline@mkMsg1@ to construct \lstinline@msg1@ as an output,
given its other parameters as inputs:
\begin{lstlisting}
predicate mkMsg1(msg1:item,U:item,S:item,kr:bytes,cert:bytes,
                   n:bytes,id1:string,t1:string,b1:item) :-
  isService(S,uri,ac,subj),
  isX509Cert(cert,kr,subj,"rsasha1",ek),
  isUserTokenKey(utok,U,n,t1,sk),
  mkEncryptedData(eutok,utok,ek),
  mkSignature(sig1,"hmacsha1",sk,
    <list>
      <Body>b1</>
      <To>uri</>
      <Action>ac</>
      <MessageId>id1</>
      <Created>t1</>
      eutok</>),
  env1(msg1,uri,ac,id1,t1,eutok,sig1,b1).
\end{lstlisting}

On the server side, with server certificate \lstinline@cert@, associated private key \lstinline@sx@,
and expected user \lstinline@U@, we use a predicate \lstinline@isMsg1@ to check
the input \lstinline@msg1@ and produce \lstinline@S@, \lstinline@id1@,
\lstinline@t1@, and \lstinline@b1@ as outputs:
\begin{lstlisting}
predicate isMsg1(msg1:item,U:item,sx:bytes,cert:bytes,S:item,
                 id1:string,t1:string,b1:item) :-
  env1(msg1,uri,ac,id1,t1,eutok,sig1,b1),
  isService(S,uri,ac,subj),
  isEncryptedData(eutok,utok,sx),
  isUserTokenKey(utok,U,n,t1,sk),
  isSignature(sig1,"hmacsha1",sk,
    <list>
      <Body>b1</>
      <To>uri</>
      <Action>ac</>
      <MessageId>id1</>
      <Created>t1</>
      eutok</>).
\end{lstlisting}

\subsection{Building and Checking Message~2}
\label{sec:message2}

Our goal \lstinline{C2} is to reach agreement on the data
\begin{quote}
\lstinline@(U,S,id1,t1,b1,id2,t2,b2)@
\end{quote}
where
\begin{quote}
\lstinline@U@=\lstinline@<User><Username>u</><Password>pwd</></>@\\
\lstinline@S@=\lstinline@<Service><To>uri</><Action>ac</><Subject>subj</></>@
\end{quote}
after successful receipt of Message~2,
having already agreed on
\begin{quote}
\lstinline@(U,S,id1,t1,b1)@
\end{quote}
after receipt of Message~1.

A simple implementation is to make sure that the client's choice of
\lstinline@id1@ in Message~1 is fresh and unpredictable, to include
\lstinline@<relatesTo>id1</>@ in Message~2, and to embed this element
in the signature to achieve correlation with the data sent in
Message~1.
In more detail, Message~2 includes a certificate for~\lstinline@S@
(that is, one with subject name \lstinline@subj@) and a signature
token, signing (elements containing) \lstinline@id1@, \lstinline@id2@,
\lstinline@t2@, and \lstinline@b2@ and signed using the private key
associated with \lstinline$S$'s certificate.
The structure of the second envelope is defined as follows:
\adg{``From'' doesn't need to be signed}%
\begin{lstlisting}
predicate env2(msg2:item,uri:item,id1:string,id2:string,
               t2:string,cert:bytes,sig2:item,b2:item) :-
  msg2 =
    <Envelope>
      <Header>
        <From>uri</>
        <RelatesTo>id1</>
        <MessageId>id2</>
        <Security>
          <Timestamp><Created>t2</></>
          <BinarySecurityToken>base64(cert)</>
          sig2</></>
      <Body>b2</></>.
\end{lstlisting}

A server uses the predicate \lstinline@mkMsg2@ to construct \lstinline@msg2@ as an output,
given its other parameters as inputs (including the signing key):
\begin{lstlisting}
predicate mkMsg2(msg2:item,sx:bytes,cert:bytes,S:item,
                 id1:string,id2:string,t2:string,b2:item):-
  isService(S,uri,ac,subj),
  mkSignature(sig2,"rsasha1",sx,
    <list>
      <Body>b2</>
      <RelatesTo>id1</>
      <MessageId>id2</>
      <Created>t2</></>),
  env2(msg2,uri,id1,id2,t2,cert,sig2,b2).
\end{lstlisting}

A client, given the CA's public key \lstinline@kr@, and awaiting a
response from \lstinline@S@ to a message with unique identifier \lstinline@id1@,
uses the predicate \lstinline@isMsg2@ to check its input \lstinline@msg2@, and produce
data \lstinline@id2@, \lstinline@t2@, and \lstinline@b2@ as outputs.
\begin{lstlisting}
predicate isMsg2(msg2:item,S:item,kr:bytes,
                 id1:string,id2:string,t2:string,b2:item) :-
  env2(msg2,uri,id1,id2,t2,cert,sig2,b2),
  isService(S,uri,ac,subj),
  isX509Cert(cert,kr,subj,"rsasha1",k),
  isSignature(sig2,"rsasha1",k,
    <list>
      <Body>b2</>
      <RelatesTo>id1</>
      <MessageId>id2</>
      <Created>t2</></>).
\end{lstlisting}

\section{Processes and Assertions in TulaFale}
\label{sec:proc-assertions-TulaFale}

A TulaFale script defines a system to be a collection of concurrent
processes that may compute internally, using terms and predicates, and
may also communicate by exchanging terms on named channels.
The top-level process defined by a TulaFale script represents the
behaviour of the principals making up the system---some clients and
servers in our example.
The attacker is modelled as an arbitrary process running
alongside the system defined by the script, interacting with it via
the public channels.
The style of modelling cryptographic protocols, with an explicit given process
representing the system and an implicit arbitrary process representing the
attacker, originates with the spi calculus~\cite{AG99:spi}.
We refer to the principals coded explicitly as processes in the script
as being \emph{compliant}, in the sense they are constrained to follow
the protocol being modelled, as opposed to the non-compliant
principals represented implicitly by the attacker process, who are not
so constrained.

A TulaFale script consists of a sequence of declarations.
We have seen already in Sections~\ref{sec:tulafale-basics}
and~\ref{sec:modelling-envelopes} many examples of Prolog-style declarations
of clauses defining named predicates.
This section describe three further kinds of declaration---for
channels, correspondence assertions, and processes.
Section~\ref{sec:modelling-protocol} illustrate their usage in a
script that models the system of Section~\ref{sec:example}.

We describe TulaFale syntax in terms of several metavariables or nonterminals:
{\lstinline{sort}}, {\lstinline{term}}, and {\lstinline{form}} range over
the sorts, algebraic terms, and logical formulas, respectively,
as introduced in Section~\ref{sec:tulafale-basics};
and {\lstinline{ide}} ranges over alphanumeric identifiers, used to name
variables, predicates, channels, processes, and correspondence assertions.

A declaration \lstinline{channel ide(sort$_1$, $\dots$, sort$_n$)} introduces a
channel, named {\lstinline{ide}}, for exchanging $n$-tuples of terms with sorts
\lstinline{sort}$_1$, $\dots$, \lstinline{sort}$_n$.
As in the asynchronous {\picalc}, channels are named, unordered queues of
messages.
By default, each channel is public, that is, the attacker may input or output
messages on the channel.
The declaration may be preceded by the \lstinline{private} keyword to prevent
the attacker accessing the channel.
Typically, SOAP channels are public, but channels used to represent
shared secrets, such as passwords, are private.

In TulaFale, as in some forms of the spi calculus, we embed correspondence
assertions in our process language in order to state certain security
properties enjoyed by compliant principals.

A declaration \lstinline{correspondence ide(sort$_1$, $\dots$, sort$_n$)}
introduces a label, {\lstinline{ide}}, for events represented by $n$-tuples
of terms with sorts \lstinline{sort}$_1$, $\dots$, \lstinline{sort}$_n$.
Each event is either a begin-event or an end-event; typically, a
begin-event records an initiation of a session, and an end-event
records the satisfactory completion of a session, from the compliant
principals' viewpoint. 
The process language includes constructs for logging begin- and end-events.
The attacker cannot observe or generate events.
We use correspondences to formalize the
properties (\lstinline{C1}) and (\lstinline{C2}) of
Section~\ref{sec:example}.
The declaration of a correspondence on {\lstinline{ide}} specifies a security assertion:
that in any run of the explicit system composed with an arbitrary implicit
attacker, every end-event labelled {\lstinline{ide}} logged by the system corresponds to
a previous begin-event logged by the system, also labelled {\lstinline{ide}}, and with
the same tuple of data.
We name this property \emph{robust safety}, following Gordon and
Jeffrey~\cite{GJ03:AuthenticityByTyping}.
The implication of robust safety is that two compliant processes have
reached agreement on the data, which typically include the contents of
a sequence of one or more messages.

A declaration \lstinline{process ide(ide$_1$:sort$_1$, $\dots$, ide$_n$:sort$_n$) = proc}
defines a parametric process, with body the process
{\lstinline{proc}}, named {\lstinline{ide}}, whose parameters
{\lstinline{ide}$_1$, $\dots$, \lstinline{ide}$_n$} have sorts
\lstinline{sort}$_1$, $\dots$, \lstinline{sort}$_n$, respectively.

Next, we describe the various kinds of TulaFale process.
\begin{itemize}
\item
A process \lstinline{out ide(tm$_1$, $\dots$, tm$_n$); proc}
sends the tuple (\lstinline{tm$_1$}, $\dots$, \lstinline{tm$_n$}) 
on the {\lstinline{ide}} channel,
then runs {\lstinline{proc}}.
\item
A process \lstinline{in ide(ide$_1$, $\dots$, ide$_n$); proc}
blocks awaiting a tuple (\lstinline{tm$_1$}, $\dots$, \lstinline{tm$_n$}) on the {\lstinline{ide}}
channel; if one arrives, the process behaves as {\lstinline{proc}},
with its parameters \lstinline{ide$_1$}, $\dots$, \lstinline{ide$_n$} bound to
\lstinline{tm$_1$}, $\dots$, \lstinline{tm$_n$}, respectively.
\item
A process \lstinline{new ide:bytes; proc} binds the variable {\lstinline{ide}} to a fresh
byte array, to model cryptographic key or nonce generation, for
instance, then runs {\lstinline{proc}}.
Similarly, a process \lstinline{new ide:string; proc} binds the variable
{\lstinline{ide}} to a fresh string, to model password generation, for instance, then
runs as {\lstinline{proc}}.
\item
A process \lstinline{proc$_1$ | proc$_2$} is a parallel composition of
subprocesses {\lstinline{proc$_1$}} and {\lstinline{proc$_2$}}; they run in parallel, and may
communicate on shared channels.
\item
A process \lstinline{! proc} is a parallel composition of
an unbounded array of replicas of the process {\lstinline{proc}}.
\item
The process \lstinline{0} does nothing.
\item
A process \lstinline{let ide = tm; proc}
binds the term {\lstinline{tm}} to the variable {\lstinline{ide}},
then runs {\lstinline{proc}}.
\item
A process \lstinline{filter form -> ide$_1$, $\dots$, ide$_n$; proc}
binds terms \lstinline{tm$_1$}, $\dots$, \lstinline{tm$_n$}
to the variables \lstinline{ide$_1$}, $\dots$, \lstinline{ide$_n$}
such that the formula {\lstinline{form}} holds, then runs \lstinline{proc}.
(The terms \lstinline{tm$_1$}, $\dots$, \lstinline{tm$_n$} are computed by pattern-matching, as
described in a previous
paper~\cite{BFG04:SemanticsForWebServicesAuthentication}.)
\item
A process \lstinline{ide(tm$_1$, $\dots$, tm$_n$)},
where \lstinline{ide} corresponds to a declaration 
\lstinline{process} \linebreak[2] \lstinline{ide(ide$_1$:sort$_1$, $\dots$ , ide$_n$:sort$_n$) = proc}
binds the terms \lstinline{tm$_1$}, $\dots$, \lstinline{tm$_n$}
to the variables \lstinline{ide$_1$}, $\dots$, \lstinline{ide$_n$},
then runs {\lstinline{proc}}.
\item
A process \lstinline{begin ide(tm$_1$, $\dots$, tm$_n$); proc} logs a
begin-event labelled with {\lstinline{ide}} and the tuple
(\lstinline{tm$_1$}, $\dots$, \lstinline{tm$_n$}), then runs {\lstinline{proc}}.
\item
A process  \lstinline{end ide(tm$_1$, $\dots$, tm$_n$); proc} logs an
end-event labelled with {\lstinline{ide}} and the tuple
(\lstinline{tm$_1$}, $\dots$, \lstinline{tm$_n$}), then runs {\lstinline{proc}}.
\item Finally, the process \lstinline{done} logs a done-event.  
  (We typically mark the successful completion of the whole protocol
  with \lstinline{done}.  Checking for the reachability of the
  done-event is then a basic check of the functionality of the
  protocol, that it may run to completion.)
\end{itemize}

The main goal of the TulaFale tool is to prove or refute robust safety
for all the correspondences declared in a script.
Robust safety may be proved by a range of techniques; the first paper on
TulaFale~\cite{BFG04:SemanticsForWebServicesAuthentication} uses manually
developed proofs of behavioural equivalence.
Instead, our TulaFale tool translates scripts into the applied {\picalc}, and
then runs Blanchet's resolution-based protocol verifier; the translation is
essentially the same as originally
described~\cite{BFG04:SemanticsForWebServicesAuthentication}.
ProVerif (hence TulaFale) can also check secrecy assertions, but we omit the details here.
In addition, TulaFale includes a sort-checker and a simulator, both of
which help catch basic errors during the development of scripts.
For example, partly relying on the translation, TulaFale can show the
reachability of \lstinline@done@ processes, which is useful for
verifying that protocols may actually run to completion.

\section{Modelling and Verifying our Protocol}
\label{sec:modelling-protocol}

Relying on the predicates given in
Section~\ref{sec:modelling-envelopes}, we now define the processes
modelling our sample system.

\subsection{Modelling the System with TulaFale Processes}
\label{sec:system-modelling}

In our example, a public channel \lstinline@publish@ gives the attacker access to
the certificates and public keys of the CA and two servers,
named \lstinline@"BobsPetShop"@ and \lstinline@"ChasMarket"@.
Another channel \lstinline@soap@ is for exchanging all SOAP messages; it is
public to allow the attacker to read and write any SOAP message.
\begin{lstlisting}
channel publish(item).
channel soap(item).
\end{lstlisting}

The following is the top-level process, representing the behaviour of all
compliant principals.
\begin{lstlisting}[linerange=222-230]
new  sr:bytes; let kr = pk(sr);
new sx1:bytes; let cert1 = x509(sr,"BobsPetShop","rsasha1",pk(sx1));
new sx2:bytes; let cert2 = x509(sr,"ChasMarket","rsasha1",pk(sx2));
out publish(base64(kr));
out publish(base64(cert1));
out publish(base64(cert2));
( !MkUser(kr) | !MkService(sx1,cert1) | !MkService(sx2,cert2) |
  (!in anyUser(U); Client(kr,U)) |
  (!in anyService(sx,cert,S); Server(sx,cert,S)) )
\end{lstlisting}

The process begins by generating the private and public keys, \lstinline@sr@ and \lstinline@kr@, of the CA.
It generates the private keys, \lstinline@sx1@ and \lstinline@sx@2, of the two
servers, plus their certificates, \lstinline@cert1@ and \lstinline@cert2@.
Then it outputs the public data \lstinline@kr@, \lstinline@cert1@, \lstinline@cert2@ to the attacker.
After this initialization, the system behaves as the following parallel
composition of five processes.
\begin{lstlisting}
  !MkUser(kr) | !MkService(sx1,cert1) | !MkService(sx2,cert2) |
  (!in anyUser(U); Client(kr,U)) |
  (!in anyService(sx,cert,S); Server(sx,cert,S)
\end{lstlisting}
As explained earlier, the \lstinline@!@ symbol represents unbounded
replication; each of these processes may execute arbitrarily many times.
The first process allows the opponent to generate fresh
username/password combinations that are
shared between all clients and servers.
The second and third allow the opponent to generate fresh services
with subject names \lstinline@"BobsPetShop"@ and \lstinline@"ChasMarket"@, respectively.
The fourth acts as an arbitrary client \lstinline{U},
and the fifth acts as an arbitrary service \lstinline{S}.

The process \lstinline@MkUser@ inputs the name of a user from the
environment on channel \lstinline@genUser@, then creates a new password and
records the username/password combination \lstinline$U$ as a message
on private channel \lstinline@anyUser@, representing the database of
authorized users.

\begin{lstlisting}[firstline= 165,lastline= 171]
channel genUser(string).
private channel anyUser(item).
process MkUser(kr:bytes) = 
  in genUser(u);
  new pwd:string; 
  let U = <User><Username>u</><Password>pwd</></>;
  !out anyUser (U).
\end{lstlisting}

The process \lstinline{MkService} allows the attacker
to create services with the subject name of the certificate \lstinline{cert}.

\begin{lstlisting}[firstline= 176,lastline= 185]
predicate isServiceData(S:item,sx:bytes,cert:bytes) :-
  isService(S,uri,ac,x509user(cert)),
  pk(sx) = x509key(cert).

channel genService(item).
private channel anyService(bytes,bytes,item).
process MkService(sx:bytes,cert:bytes) = 
  in genService(S);
  filter isServiceData(S,sx,cert) -> ;
  !out anyService(sx,cert,S).
\end{lstlisting}

Finally, we present the processes representing clients and servers.
Our desired authentication and correlation properties are correspondence
assertions embedded within these processes.
We declare the sorts of data to be agreed by the clients and servers as follows.
\begin{lstlisting}
correspondence C1(item,item,string,string,item).
correspondence C2(item,item,string,string,item,string,string,item).
\end{lstlisting}

\adg{code edited slightly}
The process \lstinline{Client(kr:bytes,U:item)} acts as a client for the user \lstinline{U}, assuming that \lstinline{kr} is the CA's public key.
\begin{lstlisting}[firstline= 190,lastline= 201]
channel init(item,bytes,bytes,string,item).
process Client(kr:bytes,U:item) = 
  in init (S,certA,n,t1,b1);
  new id1:string;
  begin C1 (U,S,id1,t1,b1);
  filter mkMsg1(msg1,U,S,kr,certA,n,id1,t1,b1) -> msg1;

  out soap(msg1);
  in soap(msg2);
  filter isMsg2(msg2,S,kr,id1,id2,t2,b2) -> id2,t2,b2;
  end C2 (U,S,id1,t1,b1,id2,t2,b2);
  done.
\end{lstlisting}%
The process generates a globally fresh, unpredictable identifier \lstinline{id1} for
Message~1, to allow correlation of Message~2 as discussed above.
It then allows the attacker to control the rest of the data to be
sent by receiving it off the public \lstinline{init} channel.
Next, the TulaFale \lstinline{filter} operator evaluates the predicate \lstinline{mkMsg1}
to bind variable \lstinline{msg1} to Message~1.
At this point, the client marks its intention to communicate data to the server
by logging an end-event, labelled \lstinline{C1}, and then outputs the message \lstinline{msg1}.
The process then awaits a reply, \lstinline{msg2}, checks the reply with the
predicate \lstinline{isMsg2}, and if this check succeeds, ends the \lstinline{C2}
correspondence.
Finally, to mark the end of a run of the protocol, it becomes the \lstinline{done}
process---an inactive process, that does nothing, but whose reachability can
be checked, as a basic test of the protocol description.

The process \lstinline{Server(sx:bytes,cert:bytes,S:item)} represents a
service \lstinline{S}, with private key \lstinline{sx}, and certificate
\lstinline{cert}.
\begin{lstlisting}[firstline= 207,lastline= 217]
channel accept(string,string,item).
process Server(sx:bytes,cert:bytes,S:item) = 
  in soap(msg1);
  in anyUser(U);
  filter isMsg1(msg1,U,sx,cert,S,id1,t1,b1) -> id1,t1,b1;
  end C1 (U,S,id1,t1,b1);

  in accept (id2,t2,b2);
  filter mkMsg2(msg2,sx,cert,S,id1,id2,t2,b2) -> msg2;
  begin C2 (U,S,id1,t1,b1,id2,t2,b2);
  out soap(msg2).
\end{lstlisting}%
The process begins by selecting a SOAP message \lstinline{msg1} and a user \lstinline{U} off
the public \lstinline{soap} and the private \lstinline{anyUser} channels, respectively.
It filters this data with the \lstinline{isMsg1} predicate, which checks that
\lstinline{msg1} is from \lstinline{U}, and binds the variables \lstinline{S},
\lstinline{id1}, \lstinline{t1}, and \lstinline{b1}.
At this point, it asserts an end-event, labelled \lstinline{C1}, to signify apparent
agreement on this data with a client.
Next, the process inputs data from the opponent on channel \lstinline{accept},
to determine the response message.
The server process completes by using the predicate \lstinline{mkMsg2} to construct
the response \lstinline{msg2}, asserting a begin-event for the \lstinline{C2}
correspondence, and finally sending the message.

\subsection{Analysis}
\label{sec:analysis}

The TulaFale script for this example protocol consists of 200 lines specific
to the protocol, and 200 lines of library predicates.
(We have embedded essentially the whole script in this paper.)
Given the applied {\picalc} translation of this script, ProVerif shows that
our two correspondences \lstinline{C1} and \lstinline{C2} are robustly safe.
Failure of robust safety for \lstinline{C1} or \lstinline{C2} would reveal that the server
or the client has failed to authenticate Message~1, or to authenticate
Message~2 and correlate it with Message~1, respectively.
Processing is swift enough---around 25s on a 2.4GHz 1GB P4---to support easy
experimentation with variations in the protocol, specification, and threat
model.

\section{Conclusions}
\label{sec:conc}

TulaFale is a high-level language based on XML with symbolic cryptography,
clausally-defined predicates, {\picalc} processes, and correspondence
assertions.
Previous work~\cite{BFG04:SemanticsForWebServicesAuthentication} introduces a
preliminary version of TulaFale, defines its semantics via translation into
the applied {\picalc}~\cite{AbadiFournet:applied-pi-popl01}, illustrates
TulaFale via several single-message protocols, and describes hand-crafted
correctness proofs.

The original reasons for choosing to model {\wssec} protocols using the the
{\picalc}, rather than some other formal method, include the generality of the
threat model (the attacker is an unbounded, arbitrary process), the ease of
simulating executable specifications written in the {\picalc}, and the
existence of a sophisticated range of techniques for reasoning about
cryptographic protocols.

Blanchet's ProVerif~\cite{BlanchetCSFW01,BlanchetSAS02} turns out to be a
further reason for using the {\picalc} to study SOAP security.
Our TulaFale tool directly implements the translation into the applied
{\picalc} and then invokes Blanchet's verifier, to obtain fully automatic
checking of SOAP security protocols.
This checking shows no attacker expressible as a formal process
can violate particular SOAP-level authentication or secrecy properties.
Hence, we expect TulaFale will be useful for describing and checking
security aspects of web services specifications.
We have several times been surprised by vulnerabilities discovered by the
TulaFale tool and the underlying verifier.
Of course, every validation method, formal or informal, abstracts some
details of the underlying implementation, so checking with TulaFale
only partially rules out attacks on actual implementations.
Still, ongoing work is exploring how to detect vulnerabilities in web services
deployments, by extracting TulaFale scripts from XML-based configuration
data~\cite{BFG04:VerifyingPolicyBasedSecurityForWebServices}.

The request/response protocol presented here is comparable to the abstract RPC
protocols proposed in earlier work on securing web
services~\cite{GP02:ValididatingWSSecurityAbstraction}, but here we accurately
model the SOAP and {\wssec} syntax used on the wire.
Compared with the SOAP-based protocols in the article introducing the TulaFale
semantics~\cite{BFG04:SemanticsForWebServicesAuthentication}, the novelties
are the need for the client to correlate the request with the reply, and the
use of encryption to protect weak user passwords from dictionary attacks.
In future work, we intend to analyse more complex SOAP protocols, such as
WS-SecureConversation~\cite{WSSecureConversation}, for securing client-server
sessions.

Although some other process calculi manipulate
XML~\cite{BS03:Iota,GM03:ModelingDynamicWebData},
they are not intended for security applications.
We are beginning to see formal methods applied to web services specifications, such as
the TLA+ specification~\cite{JLLV04:FormalSpecificationOfAWebServicesProtocol}
of the Web Services Atomic Transaction protocol, checked with the TLC model
checker.
Still, we are aware of no other security tool for web services able to analyze
protocol descriptions for vulnerabilities to XML rewriting attacks.

\paragraph{Acknowledgement}
We thank Bruno Blanchet for making ProVerif available, and for implementing
extensions to support some features of TulaFale.
Vittorio Bertocci, Ricardo Corin, Amit Midha, and the anonymous reviewers made
useful comments on a draft of this paper.

\bibliographystyle{alpha}
\newcommand{\etalchar}[1]{$^{#1}$}

\end{document}

